\documentclass[nofootinbib,twocolumn,aps,pra]{revtex4}
\usepackage{graphicx}
\usepackage{epstopdf}

\begin{document}
\title{The superfluid phase transition in two-dimensional excitonic systems}
\author{V. Apinyan}
\author{T. K. Kope\'{c}\footnote{Tel.:  +48 71 395 4286; E-mail address: kopec@int.pan.wroc.pl.}}
\affiliation{Institute for Low Temperature and Structure Research, Polish Academy of Sciences\\
PO. Box 1410, 50-950 Wroc\l{}aw 2, Poland \\}

\begin{abstract}
We study the superfluid phase transition in the two-dimensional (2D) excitonic system. Employing the extended Falicov-Kimball model (EFKM) and considering the local quantum correlations in the system composed of conduction band electrons and valence band holes we demonstrate the existence of the excitonic insulator (EI) state in the system. We show that at very low temperatures, the particle phase stiffness in the pure-2D excitonic system, governed by the non-local cross correlations, is responsible for the vortex-antivortex binding phase-field state, known as the Berezinskii-Kosterlitz-Thouless (BKT) superfluid state. We demonstrate that the existence of excitonic insulator phase is a necessary prerequisite, leading to quasi-long-range order in the 2D excitonic system.
\end{abstract}

\maketitle

\section{\label{sec:Section_1} Introduction}
%
The interesting quasiparticles, excitons, play the fundamental role in the physics related to the recent revelations on the excitonic insulator (EI) state, \cite{Jerome,Phan,Zenker,Seki} excitonic Bose-Einstein condensation (BEC) \cite{Zenker, Seki, Lozovik_1, Lozovik_2} and the excitonic superfluidity, \cite{Hohenberg_1, Chang, Eisenstein, Clade, Julian, Chung, Kruger}. In analogy with Cooper pair condensate \cite{Bardeen}, one can naturally expect that the electron-hole pairs (excitons), being the bosons with neutral charges, should eventually undergo BEC or the Bose-superfluid Berezinskii-Kosterlitz-Thouless (BKT) transition \cite{Berezinskii, Kosterlitz} (in the case of negligible quantum dissipations) at the very low temperatures. The first one is typical for the three dimensional (3D) system, where it is the dominant phase transition, whereas, the second type of transition is typical for two-dimensional (2D) systems, where the long-range order is absent \cite{Mermin,Hohenberg_2}. As in the case of the usual Bose gases, there should be a relation between these two excitonic phase transitions as it is reported recently \cite{Julian}. It is well known that the BEC in the interacting uniform systems occurs only for D $>$ 2. However, the absence of BEC does not necessarily imply the lack of a superfluid phase transition in D = 2, assuming that the well-defined conditions are satisfied by the system \cite{Hohenberg_1, Chung}. In sharp contrast to the 3D case, interactions in 2D Bose system cannot be treated as a minor correction to the BEC picture of the ideal Bose gas and they qualitatively change the behavior of the system. In fact, the excitonic superfluidity requires the interacting Bose gas, \cite{Landau} on the contrary, BEC does not. The BEC is always superfluid and the existence of the critical Landau velocity \cite{Landau} in a Bose-Einstein condensed gas is well known \cite{Hadzibabic}. On the other hand, the ideal Bose gas could exhibit the phase transition to the BEC state being absolutely not superfluid. 

Despite the continuous attempts, using the ultra-high quality materials, to observe superfluidity
in bilayer electron-hole systems, such as the double wells in the GaAs-AlGaAs heterostructures \cite{Sivan, Croxall, Seamons} and graphene bilayers (in the case of graphene, barriers as thin as 1 nm), the superfluid phase has not been observed yet, except of quantum Hall regime in the presence of the external magnetic field, where the physics is quite
different \cite{Eisenstein}. Accordingly, it may seem that the electron-hole superfluidity at the vanishing external magnetic field will never occur in a solid state system, however, it has been shown \cite{Nelson_1} that a double bilayer graphene system, separated by barrier of thickness $1$ nm, should generate an excitonic superfluid at experimentally attainable temperatures, and in the case of the absence of the external magnetic field. One of the key reasons why the excitonic superfluidity is so hard to observe experimentally in 2D case is in fact related to the dominant role of quantum fluctuations at low dimensions and at low temperatures, when the very large zero-point oscillations are present. This peculiarity is due to the absence of any real heavy particle in the electron-hole (e-h) system.

In the present paper, we address the role of the particle phase coupling in the purely 2D in-plane interacting excitonic system. We explore the low-temperature quantum collective behavior of the excitons and we extend the theoretical works mentioned above, by showing that the formation of the excitonic superfluid state is governed by the non-local cross correlations between nearest neighbors (n.n.) excitonic pairs in contrast to the formation of the EI state, where the local on-site correlations are important. We derive the BKT transition lines, and we discuss the values of the physical parameters entering in the system. 
%
\section{\label{sec:Section_2} The model}
%
For the study of the EI state and the excitonic superfluidity in 2D, we have chosen the two-band extended Falicov-Kimball model (EFKM), \cite{Zenker, Seki} due to its large applicability for treatment of the electronic correlations. The Hamiltonian of the EFKM model is given by
\begin{eqnarray}
&&{\cal{H}}=-t_{c}\sum_{\left\langle {\bf{r}},{\bf{r}}' \right\rangle}\left[\bar{c}({{\bf{r}}})c({{\bf{r}}}')+h.c.\right]-\bar{\mu}\sum_{{\bf{r}}}n({\bf{r}})-
\nonumber\\
&&-t_{f}\sum_{\left\langle {\bf{r}},{\bf{r}}' \right\rangle}\left[\bar{f}({{\bf{r}}})f({{\bf{r}}}')+h.c.\right]+\frac{\epsilon_{c}-\epsilon_{f}}{2}\sum_{{\bf{r}}}\tilde{n}({\bf{r}})+
\nonumber\\
&&+U\sum_{{\bf{r}}}\frac{1}{4}\left[n^{2}({\bf{r}})-\tilde{n}^{2}({\bf{r}})\right].
\label{Equation_1}
\end{eqnarray}
Here $\bar{f}({{\bf{r}}})$ ($\bar{c}({{\bf{r}}})$) creates an $f$ ($c$) electron at the lattice position ${\bf{r}}$, the summation $\left\langle {\bf{r}}, {\bf{r}}' \right\rangle$ runs over pairs of n.n. sites of 2D square lattice, . The density type short hand notations are introduced $n({\bf{r}})=n_{c}({\bf{r}})+n_{f}({\bf{r}})$ and $\tilde{n}({\bf{r}})=n_{c}({\bf{r}})-n_{f}({\bf{r}})$. Next, $t_{c}$ is the hopping amplitude for $c$-electrons and $\epsilon_{c}$ is the corresponding on-site energy level. Similarly, $t_{f}$ is the hopping amplitude for $f$-electrons and $\epsilon_{f}$ is the on-site energy level for $f$-orbital.
The sign of the product $t_{c}t_{f}$ determines the type of semiconductor, for $t_{c}t_{f}<0$ ($t_{c}t_{f}>0$) we have a direct (indirect) band gap semiconductor. The on-site (local) Coulomb interaction $U$ in the last term of the Hamiltonian in Eq.(\ref{Equation_1}) plays the coupling role between the electrons in the $f$ and $c$ sub-systems. The chemical potential $\bar{\mu}$ is $\bar{\mu}=\mu-\bar{\epsilon}$, where $\bar{\epsilon}=\left(\epsilon_{c}+\epsilon_{f}\right)/2$. We will use $t_{c}=1$ as the unit of energy and we fix the band parameter values $\epsilon_{c}=0$ and $\epsilon_{f}=-1$. For the $f$-band hopping amplitude $t_{f}$ we consider the values $t_{f}=-0.3$ and $t_{f}=-0.1$. Throughout the paper, we set $k_{B}=1$ and $\hbar=1$ and lattice constant $a=1$.
%
\section{\label{sec:Section_3} The EI state discussion}
%
The Hamiltonian in Eq.(\ref{Equation_1}) is containing two separate quadratic terms and is suitable for decoupling by functional path integration method \cite{Negele}. We use imaginary-time fermionic path integral techniques, and we introduce the fermionic Grassmann variables ${f}({{\bf{r}}}\tau)$ and ${c}({{\bf{r}}}\tau)$ at each site ${\bf{r}}$ and for each time $\tau$, which varies in the interval $0\leq \tau \leq\beta$, where $\beta=1/T$ with $T$ being the thermodynamic temperature. The time-dependent variables ${c}({{\bf{r}}}\tau)$ and ${f}({{\bf{r}}}\tau)$ are satisfying the anti-periodic boundary conditions ${x}({{\bf{r}}}\tau)=-{x}({{\bf{r}}}\tau+\beta)$, where $x=f$ or $c$.
After decoupling the last interaction term in the Hamiltonian in Eq.(\ref{Equation_1}) we will have for the grand canonical partition function of the system
\begin{eqnarray}
{\cal{Z}}_{\rm GC}=\int \left[{\cal{D}}\bar{c}{\cal{D}}c\right]\left[{\cal{D}}\bar{f}{\cal{D}}f\right]\left[{\cal{D}}V\right]\left[{\cal{D}}\varrho\right]e^{-{\cal{S}}[\bar{c},c,{\bar{f}},f,V,\varrho]},
\label{Equation_2}
\end{eqnarray}
where the action in the exponential is given by
\begin{eqnarray}
&&{\cal{S}}[\bar{c},c,{\bar{f}},f,V,\varrho]=\sum_{{\bf{r}}}\int^{\beta}_{0}d\tau \left[\frac{V^{2}\left({\bf{r}}\tau\right)}{U}+\frac{\varrho^{2}\left({\bf{r}}\tau\right)}{U}-\right.
\nonumber\\
&&\left.-iV\left({\bf{r}}\tau\right)n\left({\bf{r}}\tau\right)-\varrho\left({\bf{r}}\tau\right)\tilde{n}\left({\bf{r}}\tau\right)\frac{}{}\right]+\sum_{x=f,c}{\cal{S}}_{B}\left[\bar{x},x\right]+
\nonumber\\
&&+\int^{\beta}_{0}d\tau{\cal{H}}\left(\tau\right).
\nonumber\\
\label{Equation_3}
\end{eqnarray}
The new variables $V\left({\bf{r}}\tau\right)$ and $\varrho\left({\bf{r}}\tau\right)$ in the action are the decoupling fields for quadratic terms in the Hamiltonian, in Eq.(\ref{Equation_1}), proportional to $n^{2}\left({\bf{r}}\tau\right)$ and $\tilde{n}^{2}\left({\bf{r}}\tau\right)$ respectively.
Next, ${\cal{S}}_{B}[\bar{f},f]$ and ${\cal{S}}_{B}[\bar{c},c]$ are Berry actions for $f$ and $c$-electrons and they are defined as follows ${\cal{S}}_{B}[\bar{x},x]=\sum_{{\bf{r}}}\int^{\beta}_{0}d\tau \bar{x}({\bf{r}}\tau)\dot{x}({\bf{r}}\tau)$, where $\dot{x}({\bf{r}}\tau)=\partial_{\tau}x({\bf{r}}\tau)$ is the time derivative. Next, we will factorize usual electron operators $f$ and $c$ in terms of new fermionic variables $\tilde{f}$ and $\tilde{c}$ coupled to the unitary charge-carrying U(1) rotor. To this end we write the potential $V\left({\bf{r}}\tau\right)$ as the sum of a static and periodic part $V\left({\bf{r}}\tau\right)=V_{0}+\tilde{V}\left({\bf{r}}\tau\right)$. Then, for the periodic part, we introduce the $U(1)$ phase field $\varphi\left({\bf{r}}\tau\right)$ via the ``Faraday"-type relation \cite{Kopec_1}
\begin{eqnarray}
\tilde{V}\left({\bf{r}}\tau\right)=\frac{\partial{\varphi\left({\bf{r}}\tau\right)}}{\partial{\tau}}.
\label{Equation_4}
\end{eqnarray}
For the static part $V_{0}$ and $\varrho\left({\bf{r}}\tau\right)$-field, the saddle-point evaluations give $V^{s.p.}_{0}=i\frac{Un}{2}-i\bar{\mu}$ and $\varrho^{s.p.}=\frac{U\tilde{n}}{2}-\frac{\epsilon_{c}-\epsilon_{f}}{2}$. Here $n$ is the average total particle density $n=\left\langle{n}_{c}\left({\bf{r}}\tau\right)\right\rangle+\left\langle {n}_{f}\left({\bf{r}}\tau\right)\right\rangle$ (furthermore, we will fix $n=1$, corresponding to the case of half-filling \cite{Zenker,Seki}) and $\tilde{n}$ is the average of the difference of particle densities  $\tilde{n}=\left\langle \tilde{n}\left({\bf{r}}\tau\right)\right\rangle$. Then the partition function of the system becomes
\begin{eqnarray}
{\cal{Z}}_{\rm GC}=\int \left[{\cal{D}}\bar{c}{\cal{D}}c\right]\left[{\cal{D}}\bar{f}{\cal{D}}f\right]\left[{\cal{D}}\varphi\right]e^{-{\cal{S}}[\bar{c},c,{\bar{f}},f,\varphi]}
\label{Equation_5}
\end{eqnarray}
and the total action in Eq.(\ref{Equation_3}) reduces to 
\begin{eqnarray}
&&{\cal{S}}[\bar{c},c,{\bar{f}},f,\varphi]={\cal{S}}_{\rm eff}\left[\varphi\right]+{\cal{S}}_{B}\left[\bar{c},c\right]+{\cal{S}}_{B}\left[\bar{f},f\right]-
\nonumber\\
&&-t_{c}\sum_{\left\langle{\bf{r}}, {\bf{r}}' \right\rangle}\int^{\beta}_{0}d\tau \left[\bar{c}({{\bf{r}}}\tau)c({{\bf{r}}}'\tau)+h.c.\right]-
\nonumber\\
&&-{t}_{f}\sum_{\left\langle {\bf{r}}, {\bf{r}}' \right\rangle}\int^{\beta}_{0}d\tau \left[\bar{f}({{\bf{r}}}\tau)f({{\bf{r}}}'\tau)+h.c.\right]+
\nonumber\\
&&\ +\sum_{{\bf{r}}}\int^{\beta}_{0}d\tau \left[{\mu}_{n}n({\bf{r}}\tau)+\mu_{\tilde{n}}\tilde{n}({\bf{r}}\tau)\right].
\label{Equation_6}
\nonumber\\
\end{eqnarray}
Here 
\begin{eqnarray}
{\cal{S}}_{\rm eff}[\varphi]=\sum_{{\bf{r}}}\int^{\beta}_{0}&&d\tau\left[\frac{\dot{\varphi}^{2}({\bf{r}}\tau)}{U}-\frac{2\bar{\mu}}{iU}\dot{\varphi}({\bf{r}}\tau)\right.-
\nonumber\\
&&\left.-i\dot{\varphi}({\bf{r}}\tau)n({\bf{r}}\tau)\right]
\label{Equation_7}
\end{eqnarray}
is the phase-only action, which contains fluctuating imaginary term $i\dot{\varphi}({\bf{r}}\tau)n({\bf{r}}\tau)$. The chemical potentials ${\mu}_{n}$ and $\mu_{\tilde{n}}$ are defined as $\mu_{n}=\frac{Un}{2}-\bar{\mu}$ and $\mu_{\tilde{n}}=\frac{\epsilon_{c}-\epsilon_{f}}{2}-\frac{U\tilde{n}}{2}$.

Next, we perform the local gauge transformation to new fermionic variables $\tilde{f}({\bf{r}}\tau)$ and $\tilde{c}({\bf{r}}\tau)$. For the electrons of $f$ and $c$ orbitals, the U$(1)$ gauge transformation could be written as
\begin{eqnarray}
\left[
\begin{array}{cc}
x({\bf{r}}\tau) \\
\bar{x}({\bf{r}}\tau)
\end{array}
\right]=\hat{{\cal{U}}}(\varphi)\cdot\left[
\begin{array}{cc}
\tilde{x}({\bf{r}}\tau) \\
\bar{\tilde{x}}({\bf{r}}\tau)
\end{array}
\right],
\label{Equation_8}
\end{eqnarray} 
where $\hat{\cal{U}}(\varphi)$ is the U(1) transformation matrix $\hat{\cal{U}}(\varphi)=\hat{I}\cdot\cos\varphi({\bf{r}}\tau)+i\hat{\sigma}_{z}\cdot\sin\varphi({\bf{r}}\tau)$ with the unit matrix $\hat{I}$ and $\hat{\sigma}_{z}$ being the Pauli matrix. \
%
\begin{figure}[!ht]
\includegraphics[width=250px,height=250px]{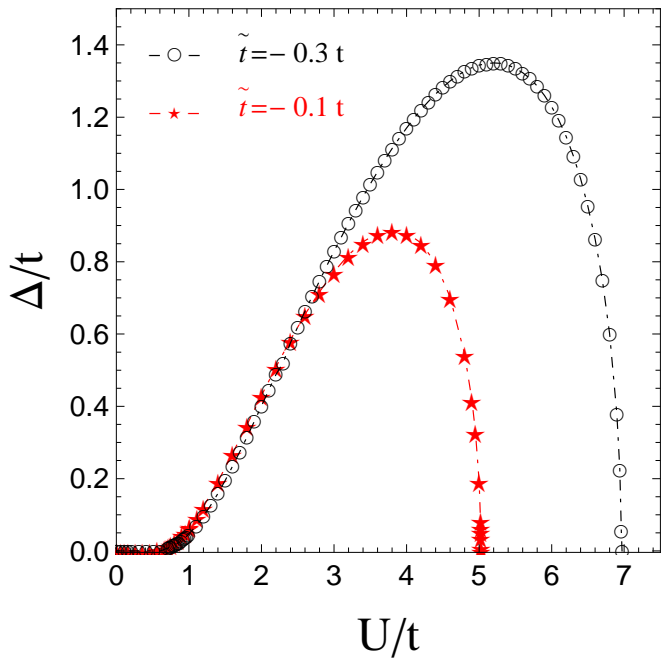} 
{\caption{\label{fig:Fig_1} The excitonic order parameter $\Delta$ as a function of Coulomb interaction parameter $U$ normalized to the hopping amplitude $t$.} }
\end{figure}
%

After those transformations we can write the total action of the system in the Fourier-space representation introducing the vector-space notations and, furthermore, we will derive gap equation for the excitonic order parameter $\Delta$. The effective phase averaged action of the system in the Fourier space takes the following form
\begin{eqnarray}
&&{\cal{S}}_{\rm eff}\left[\tilde{\bar{c}},\tilde{c},\tilde{\bar{f}},\tilde{f}\right]
=\frac{1}{\beta{N}}\sum_{{\bf{k}}\nu_{n}}\left[\bar{\tilde{c}}_{\bf{k}}(\nu_{n}),\bar{\tilde{f}}_{\bf{k}}({\nu_{n}})\right]\times
\nonumber\\
&&\times{\cal{G}}^{-1}({\bf{k}},\nu_{n})\left[\begin{array}{cc}
{\tilde{c}}_{\bf{k}}(\nu_{n})\\
{\tilde{f}}_{\bf{k}}(\nu_{n})
\end{array}
\right].
\label{Equation_9}
\end{eqnarray}
Here ${\cal{G}}^{-1}({\bf{k}},\nu_{n})$ is the inverse of the Green function matrix, given by
\begin{eqnarray}
{\cal{G}}^{-1}({\bf{k}},\nu_{n})=
\left(
\begin{array}{cc}
{E}^{\tilde{c}}_{{\bf{k}}}(\nu_{n})
 & -\bar{\Delta}  \\
-\Delta & {E}^{\tilde{f}}_{{\bf{k}}}(\nu_{n})
\end{array}
\right),
\nonumber\\
\ \ \ \ 
\label{Equation_10}
\end{eqnarray}
where single-particle Bogoliubov quasienergies ${E}^{\tilde{f}}_{{\bf{k}}}(\nu_{n})$ and ${E}^{\tilde{c}}_{{\bf{k}}}(\nu_{n})$ are given as ${E}^{\tilde{c}}_{{\bf{k}}}(\nu_{n})=\bar{\epsilon}_{\tilde{c}}-i\nu_{n}-{t}_{{\bf{k}}}$, ${E}^{\tilde{f}}_{{\bf{k}}}(\nu_{n})=\bar{\epsilon}_{\tilde{f}}-i\nu_{n}-\tilde{t}_{{\bf{k}}}$. Next, ${t}_{{\bf{k}}}$ and $\tilde{t}_{{\bf{k}}}$ are band-renormalized hopping amplitudes ${t}_{{\bf{k}}}=2t{\mathrm{g}}_{B}\gamma_{{\bf{k}}}$ and $\tilde{t}_{{\bf{k}}}=2\tilde{t}{\mathrm{g}}_{B}\gamma_{{\bf{k}}}$, where ${\mathrm{g}}_{B}$ is the bandwidth renormalization factor 
\begin{eqnarray}
\mathrm{g}_{B}=\left.\left\langle e^{-i[\varphi({{\bf{r}}}\tau)-\varphi({{\bf{r}}}'\tau)]} \right\rangle\right|_{|{\bf{r}}-{\bf{r}}'|={{d}}}
\label{Equation_11}
\end{eqnarray}
and $\gamma_{{\bf{k}}}$ is the 2D lattice dispersion $\gamma_{{\bf{k}}}=\cos(k_{x}d_{x})+\cos(k_{y}d_{y})$, with $d_{\alpha}$ ($\alpha=x,y$), being the components of the lattice spacing vector ${\bf{d}}={\bf{r}}-{\bf{r}}'$ with ${\bf{r}}$ and ${\bf{r}}'$ nearest neighbors site positions. For the simple, square plane we have $d_{\alpha}\equiv a$.  
The calculation of ${\mathrm{g}_{B}}$ within the self-consistent harmonic approximation (SCHA) \cite{Simanek, Wood_Stroud} is given in the Section \ref{sec:Section_5}. The quasiparticle energies $\bar{\epsilon}_{\tilde{f}}$ and $\bar{\epsilon}_{\tilde{c}}$ are of Hartree-type and they are defined in the theory by relation $\bar{\epsilon}_{\tilde{x}}=\epsilon_{{x}}-\mu+Un_{\tilde{y}}+i\left\langle\dot{\varphi}({{\bf{r}}}\tau)\right\rangle$, where $\tilde{y}$ means orbital, opposite to $\tilde{x}$.

The EI low-temperature phase is characterized by local excitonic order parameter $\Delta=U\left\langle \bar{\tilde{c}}({\bf{r}}\tau)\tilde{f}({\bf{r}}\tau)\right\rangle$. The EI state develops from local on-site electron-hole correlations. The expectation value, given in the expression of local EI order parameter, could be calculated in the frame of path integral method \cite{Negele} as well as the fermion density averages of the respective band levels $n_{\tilde{x}}=\left\langle\bar{\tilde{x}}({\bf{r}}\tau)\tilde{x}({\bf{r}}\tau)\right\rangle$. We get a set of coupled SC equations for the EI order parameter $\Delta$, single-particle fermion densities $n_{\tilde{x}}$ and EI chemical potential $\mu$
\begin{eqnarray}
&&\frac{1}{N}\sum_{{\bf{k}}}\left[f(E^{+}_{{\bf{k}}})+f(E^{-}_{{\bf{k}}})\right]=1,
\label{Equation_12} 
\newline\\
&&\tilde{n}=\frac{1}{N}\sum_{{\bf{k}}}\xi_{{\bf{k}}}\cdot\frac{f(E^{+}_{{\bf{k}}})-f(E^{-}_{{\bf{k}}})}{\sqrt{\xi^{2}_{{\bf{k}}}+4\Delta^{2}}},
\label{Equation_13} 
\newline\\
&&\Delta=-\frac{U\Delta}{N}\sum_{{\bf{k}}}\frac{f(E^{+}_{{\bf{k}}})-f(E^{-}_{{\bf{k}}})}{\sqrt{\xi^{2}_{{\bf{k}}}+4\Delta^{2}}}.
\label{Equation_14}  
\end{eqnarray}
Here $\xi_{{\bf{k}}}=-{t}_{{\bf{k}}}+\bar{\epsilon}_{\tilde{c}}+\tilde{t}_{{\bf{k}}}-\bar{\epsilon}_{\tilde{f}}$ is the quasiparticle dispersion and the energy parameters $E^{+}_{{\bf{k}}}$ and $E^{-}_{{\bf{k}}}$ are defined as
\begin{eqnarray}
E^{\pm}_{{\bf{k}}}=\frac{1}{2}\left(-{t}_{{\bf{k}}}+\bar{\epsilon}_{\tilde{c}}-\tilde{t}_{{\bf{k}}}+\bar{\epsilon}_{\tilde{f}}\pm{\sqrt{\xi^{2}_{{\bf{k}}}+4\Delta^{2}}}\right).
\label{Equation_15}
\end{eqnarray}
The numerical solution of the system of coupled SC equations Eqs.(\ref{Equation_12})-(\ref{Equation_14}) is performed for fixed value of the total particle density $n=n_{\tilde{f}}+n_{\tilde{c}}=1$ and ${\bf{k}}$-summations were performed with the ($100$$\times$$100$) ${\bf{k}}$-points in the First Brillouin Zone (FBZ). The finite-difference approximation method is used in numerical evaluations, which retains the fast convergence of Newton's method \cite{Powell}. The accuracy of convergence for numerical solutions is achieved with a relative error of order of $10^{-7}$. In Fig.~\ref{fig:Fig_1} the numerical results for the local excitonic order parameter $\Delta$ for the EI state are presented. Different values of $\tilde{t}$ are considered.
%
\section{\label{sec:Section_4} U(1) phase variables }
%
In this Section we integrate out the fermions in the partition function in Eq.(\ref{Equation_2}) and we obtain the bosonic total phase action of the system. We will show how the non-local (i.e. between the excitonic pairs on different lattice sites) fermionic correlations give the main contribution to the phase stiffness of the ensemble of excitons. The partition function in Eq.(\ref{Equation_2}) could be rewritten as 
%
\begin{figure}[!ht]
\includegraphics[width=250px,height=250px]{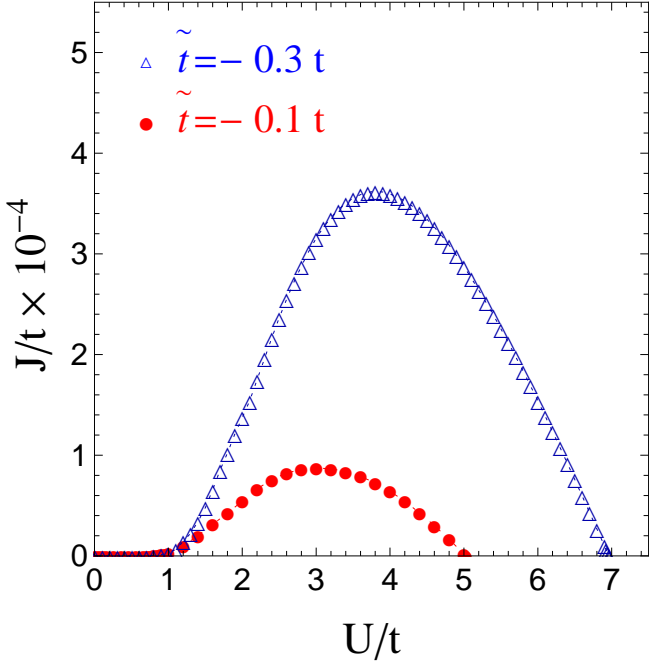} 
{\caption{\label{fig:Fig_2} The parameter $J$, as a function of the Coulomb interaction parameter $U$ normalized to the hopping amplitude $t$. In the inset, we have presented the shape of the density of states $\rho(\epsilon)$ in the case of the 2D square lattice.} }
\end{figure}
%
\begin{eqnarray}
{\cal{Z}}_{\rm GC}=\int\left[{\cal{D}}\varphi\right] e^{-{\cal{S}}_{\rm eff}[\varphi]},
\nonumber\\
\ \ \ 
\label{Equation_16}
\end{eqnarray}
where the effective phase action in the exponential is ${\cal{S}}_{\rm eff}[\varphi]=-\ln\int\left[{\cal{D}}\bar{\tilde{c}}{\cal{D}}\tilde{c}\right]\left[{\cal{D}}\bar{\tilde{f}}{\cal{D}}\tilde{f}\right] e^{-{{\cal{S}}}[\bar{\tilde{c}},{\tilde{c}},\bar{\tilde{f}},{\tilde{f}},\varphi]}$. After expanding the logarithm up to second order in the cumulant series expansion (higher terms are not considered), we find for the effective action
\begin{eqnarray}
&&{\cal{S}}_{\rm eff}[\varphi]=\tilde{{\cal{S}}}_{0}+\left\langle{{\cal{S}}}\right\rangle_{{\cal{S}}_{\rm eff}\left[{\bar{\tilde{c}},\tilde{c},\bar{\tilde{f}},\tilde{f}}\right]}-
\nonumber\\
&&-\frac{1}{2}\left[\left\langle{\cal{S}}^{2}\right\rangle_{{\cal{S}}_{\rm eff}\left[{\bar{\tilde{c}},\tilde{c},\bar{\tilde{f}},\tilde{f}}\right]}-\left\langle{\cal{S}}\right\rangle^{2}_{{\cal{S}}_{\rm eff}\left[{\bar{\tilde{c}},\tilde{c},\bar{\tilde{f}},\tilde{f}}\right]}\right].
\nonumber\\
\ \ \ 
\label{Equation_17}
\end{eqnarray}

Here ${\cal{S}}_{\rm eff}\left[\bar{\tilde{c}},\tilde{c},\bar{\tilde{f}},\tilde{f}\right]$ is the effective fermionic action given in Eq.(\ref{Equation_9}). We have replaced above $\left\langle ... \right\rangle_{\tilde{c}\tilde{f}}\rightarrow \left\langle ... \right\rangle_{{\cal{S}}_{\rm eff}\left[{\bar{\tilde{c}}},\tilde{c},\bar{\tilde{f}},\tilde{f}\right]}$, to make the calculations in a self-consistent way. We will examine the four-fermionic terms in Eq.(\ref{Equation_17}), thereby treating relevant part of non-local fermionic correlations. Therefore, the important part of the effective phase action is 
\begin{eqnarray}
{\cal{S}}_{\rm eff}[\varphi]={\cal{S}}_{0}[\varphi]+{\cal{S}}_{J}[\varphi],
\label{Equation_18}
\end{eqnarray} 
where ${\cal{S}}_{0}[\varphi]$ is the phase-only action after U(1) gauge transformation 
\begin{eqnarray}
{\cal{S}}_{0}[\varphi]=\sum_{{\bf{r}}}\int^{\beta}_{0}d\tau\left[\frac{\dot{\varphi}^{2}({\bf{r}}\tau)}{U}-\frac{2\bar{\mu}}{iU}\dot{\varphi}({\bf{r}}\tau)\right]
\label{Equation_19}
\end{eqnarray}  
and ${\cal{S}}_{J}[\varphi]=-\frac{1}{2}\left\langle{\cal{S}}^{2}\right\rangle_{{\cal{S}}_{\rm eff}\left[\bar{\tilde{c}},\tilde{c},{\bar{\tilde{f}}},\tilde{f}\right]}$. After calculating all averages in the expression of ${\cal{S}}_{J}[\varphi]$ and after not complicated evaluations we rewrite the action ${\cal{S}}_{J}[\varphi]$ in the form
\begin{eqnarray}
{\cal{S}}_{J}\left[\varphi\right]=-\frac{1}{2}\int^{\beta}_{0}d\tau \sum_{\left\langle{\bf{r}},{\bf{r}}'\right\rangle}J({\bf{r}}\tau,{\bf{r}}'\tau')\times
\nonumber\\
\times\cos{2\left[\varphi({\bf{r}}\tau)-\varphi({\bf{r}}'\tau)\right]},
\nonumber\\
\ \ \ 
\label{Equation_20}
\end{eqnarray}
where the exciton phase stiffness parameter $J$ is given by the relation
\begin{eqnarray}
J=&&\frac{\Delta^{2}t\tilde{t}}{{N^{2}}}\sum_{{\bf{k}},{\bf{k}}'}\frac{\epsilon\left({{\bf{k}}}\right)\epsilon\left({{\bf{k}}}'\right)}{{\sqrt{\xi^{2}_{{\bf{k}}}+4\Delta^{2}}}}\left[\Lambda_{1}({\bf{k}},{\bf{k}}')\tanh\left(\frac{\beta E^{+}_{{\bf{k}}}}{2}\right)-\right.
\nonumber\\
&&\left.-\Lambda_{2}({\bf{k}},{\bf{k}}')\tanh\left(\frac{\beta E^{-}_{{\bf{k}}}}{2}\right)\right].
\label{Equation_21}
\end{eqnarray}
The parameters $\Lambda_{1}({\bf{k}},{\bf{k}}')$ and $\Lambda_{2}({\bf{k}},{\bf{k}}')$ in Eq.(\ref{Equation_21}) are defined as
$\Lambda_{1}({\bf{k}},{\bf{k}}')=\left({E^{+}_{{\bf{k}}} - E^{+}_{{\bf{k}}'}}\right)^{-1}\cdot\left({E^{+}_{{\bf{k}}} - E^{-}_{{\bf{k}}'}}\right)^{-1}$ and $\Lambda_{2}({\bf{k}},{\bf{k}}')=\left({E^{-}_{{\bf{k}}} - E^{-}_{{\bf{k}}'}}\right)^{-1}\cdot\left({E^{-}_{{\bf{k}}} - E^{+}_{{\bf{k}}'}}\right)^{-1}$. The form of $J$ in Eq.(\ref{Equation_21}) indicates that the phase stiffness in the system of excitonic pairs is characterized by an energy scale proportional $\sim (\Delta t_{e}t_{h})/({t_{e}+t_{h}})$ for all the values of the Coulomb interaction parameter $U$ and it is related to the motion of the center of mass of e-h composed quasiparticle, because $(t_{e}t_{h})/(t_{e}+t_{h}) \approx (m_{e}+m_{h})^{-1}$, applying
that the exchange coupling parameter becomes proportional to the excitonic BEC-like quasi-condensate critical temperature in 2D \cite{Kagan}. In numerical evaluation of the stiffness parameter $J$, given in Eq.(\ref{Equation_21}) we transform the summations over ${\bf{k}}$ into energy integrals, by introducing density of states $\rho(\epsilon)=\frac{1}{N}\sum_{{\bf{k}}}\delta\left(\epsilon-\gamma_{{\bf{k}}}\right)$. For the tight-binding hopping matrix on the square lattice we have $\rho(\epsilon)=K\left(1-\epsilon^{2}/4\right)/\pi^{2}$, where $K(x)$ stands for the complete elliptic integral of the first kind \cite{Abramovich}. The numerical evaluations of $J$ for the case $T=0$ are shown in Fig.~\ref{fig:Fig_2}. Considering a $5$-nm GaAs coupled quantum well (QW), separated by a $4$-nm Al$_{0.33}$Ga$_{0.67}$As barrier \cite{Takahashi} with the corresponding effective electron mass m$_{e}=50.061 m_{0}$ ($m_{0}$ is the free-electron mass) and in-plane effective hole-mass of around $0.1$m$_{0}$ (according with the Luttinger parameters \cite{Takahashi}), we use the exciton binding energy \cite{Takahashi, Szymanska} value $6.7$ meV to calculate the energy scales in the model. Particularly, for the energy scale corresponding to $J$ we find for the quasi-2D GaAs/AlGaAs QW structure geometry $J\approx 0.001796 $ meV (corresponding to $U=0.0188$ eV) or, in temperature units $J\approx 20$ mK. 
%
\section{\label{sec:Section_5} BKT transition}
%
To understand the nature of the 2D excitonic low-temperature phase we describe the excitonic system far beyond the mean field description \cite{Zenker, Seki}. We will use the forme of phase-stiffness action in Eq.(\ref{Equation_20}) to derive the harmonic action in the SCHA schem. Furthermore, using the Feynmann-Kleinert (FK) method \cite{Kleinert_1, Seunghwan}, we reduce the quantum rotor model given by the action in Eq.(\ref{Equation_18}) to a simple classical Hamiltonian description. First, let's simplify the form of the phase-stiffness action in Eq.(\ref{Equation_20})
%
\begin{figure}[!ht]
\includegraphics[width=240px,height=240px]{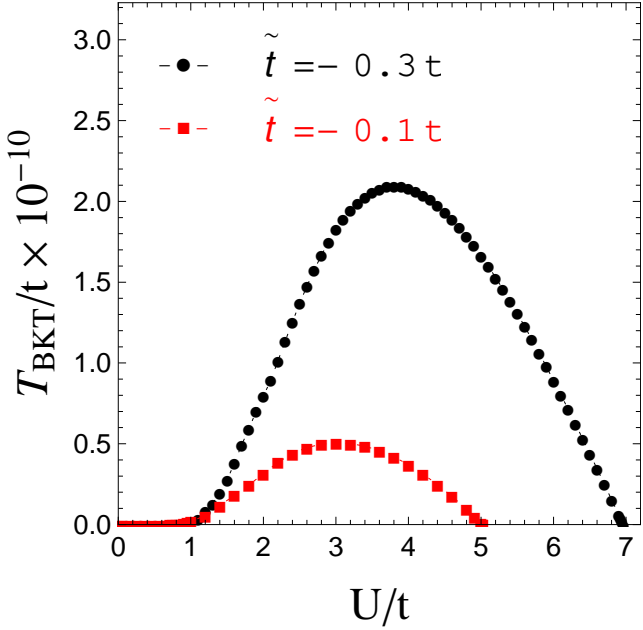} 
{\caption{\label{fig:Fig_3} Critical temperature of the BKT-superfluid phase transition in the 2D excitonic system.}}
\end{figure}
%
by transforming the linear trigonometric function inside, into a quadratic one, and by linearizing the latter one with using the standard decoupling procedure, thus, we have $\cos{2\left[\varphi({\bf{r}}\tau)-\varphi({\bf{r}}'\tau)\right]}=2\cos^{2}{\left[\varphi({\bf{r}}\tau)-\varphi({\bf{r}}'\tau)\right]}-1\approx 4\cos{\left[\varphi({\bf{r}}\tau)-\varphi({\bf{r}}'\tau)\right]}\left\langle \cos{\left[\varphi({\bf{r}}\tau)-\varphi({\bf{r}}'\tau)\right]}\right\rangle-1$. We get the linearized version of the action in Eq.(\ref{Equation_20}) ${\cal{S}}_{J}\left[\varphi\right]=-\tilde{J}\int^{\beta}_{0}d\tau \sum_{\left\langle{\bf{r}},{\bf{r}}'\right\rangle}\cos{\left[\varphi({\bf{r}}\tau)-\varphi({\bf{r}}'\tau)\right]}$, where $\tilde{J}$ now, is the renormalized effective stiffness-parameter $\tilde{J}=2J{\mathrm{g}_{B}}$. The total phase action, including also the kinetic term is
\begin{eqnarray}
&&{\cal{S}}_{\rm eff}\left[\varphi\right]=\sum_{{\bf{r}}}\int^{\beta}_{0}d\tau \frac{\dot{\varphi}^{2}({\bf{r}}\tau)}{U} +
\nonumber\\
&&+\tilde{J}\sum_{\left\langle{\bf{r}},{\bf{r}}'\right\rangle}\int^{\beta}_{0}d\tau\left\{1-\cos\left[\varphi\left({\bf{r}}\tau\right)-\varphi\left({\bf{r}}'\tau\right)\right]\right\}.
\label{Equation_22}
\end{eqnarray}
We have added a constant term to the phase stiffness action ${\cal{S}}_{J}\left[\varphi\right]$, in order to follow the standard notations \cite{Simanek, Wood_Stroud}. Next, we apply the self-consistent-harmonic-approximation (SCHA) \cite{Simanek, Wood_Stroud} to replace the action in Eq.(\ref{Equation_22}) by a trial harmonic action ${\cal{S}}_{\rm SCHA}\left[\varphi\right]$
\begin{eqnarray}
{\cal{S}}_{\rm SCHA}\left[\varphi\right]=\sum_{{\bf{r}}}\int^{\beta}_{0}d\tau \frac{\dot{\varphi}^{2}({\bf{r}}\tau)}{U}+
\nonumber\\
+\frac{\Gamma}{2}\sum_{\left\langle{\bf{r}},{\bf{r}}'\right\rangle}\int^{\beta}_{0}d\tau \left[\varphi({\bf{r}}\tau)-\varphi({\bf{r}}'\tau)\right]^{2}.
\label{Equation_23}
\end{eqnarray}
Here, $\Gamma\equiv \Gamma\left({\bf{r}}-{\bf{r}}'\right)$ is a new stiffness constant, which we will determine by using the variational approach in the theory of SCHA. We obtain the system of coupled SC equations for the bandwidth renormalization factor $\mathrm{g}_{B}$ given in Eq.(\ref{Equation_11}) and the constant $\Gamma$ in Eq.(\ref{Equation_23}) is (the case $T=0$ is considered here)
\begin{eqnarray}
&&\mathrm{g}_{B}=1-\frac{\Gamma U}{4J\mathrm{g}_{B}N}\sum_{{\bf{k}}}\frac{1-\cos\left({\bf{k}}{\bf{d}}\right)}{\sqrt{2U\Gamma f_{{\bf{k}}}}},
\nonumber\\
&&\Gamma({\bf{r}}-{\bf{r}}')=2\mathrm{g}_{B}J\cdot\exp\left[-\frac{1}{2N}\sum_{{\bf{k}}}\frac{1-\cos\left({\bf{k}}{\bf{d}}\right)}{\sqrt{2U\Gamma f_{{\bf{k}}}}}\right].
\nonumber\\
\ \ \ 
\label{Equation_24}
\end{eqnarray}

For a given $U$ and $J$, the system of SC equations in Eq.(\ref{Equation_24}) has many roots of $\mathrm{g}_{B}$. Among those solutions, only the largest one must be chosen, because it minimizes the effective action in Eq.(\ref{Equation_22}). By solving numerically the system in Eq.(\ref{Equation_24}), we got the largest root of $\mathrm{g}_{B}$ being equal to unity for all fixed values of the pair $(U,J)$, thus $\mathrm{g}_{B}\equiv 1$ in our model (this is true, if we are interesting of the low-temperature bound-vortex-antivortex type configuration, but in general $\mathrm{g}_{B}$ differs from unity in the unbinding-vortex state of the system). The partition function of the system is now
\begin{eqnarray}
{\cal{Z}}=\int \left[{\cal{D}}\varphi\right]\exp\left\{-\int^{\beta}_{0}d\tau \left[\sum_{{\bf{r}}}\frac{{\dot{\varphi}}^{2}\left({\bf{r}}\tau\right)}{U}-\right.\right.
\nonumber\\
\left.\left.-\tilde{J}\sum_{\left\langle {\bf{r}},{\bf{r}}' \right\rangle}\cos\left[\varphi({\bf{r}}\tau)-\varphi({\bf{r}}'\tau)\right]\right]\right\}.
\label{Equation_25}
\end{eqnarray}
Now, we will derive from Eq.(\ref{Equation_25}) the effective classical Hamiltonian analogue to that of 2D $X-Y$ model. We use the FK method for the classical action in Eq.(\ref{Equation_25}) combined with the SCHA method. Following \cite{Kleinert_1}, we decompose the phase variables into static and periodic parts $\varphi\left({\bf{r}}\tau\right)=\varphi_{0}\left({\bf{r}}\right)+\sum^{\infty}_{n=1}\left[\varphi_{n}\left({\bf{r}}\right)e^{i\omega_{n}\tau}+c.c.\right]$. Then, using the FK method, we choose a trial partition function ${\cal{Z}}'$ corresponding to ${\cal{Z}}$ in Eq.(\ref{Equation_25}), in which the potential energy is approximated by a Gaussian component for the part of the action with $n\neq 0$. We employ the extremal principle to find ${\cal{Z}}'$ and we have ${\cal{Z}}'=\int\prod_{{\bf{r}}}\left[{\cal{D}}\varphi\right]e^{-\beta{\cal{H}}_{\rm eff}\left[\varphi\right]}$, where the classical effective Hamiltonian ${\cal{H}}_{\rm eff}\left[\varphi\right]$ is given by
\begin{eqnarray}
&&{\cal{H}}_{\rm eff }\left[\varphi\right]=-{\cal{F}}_{\rm eff}\left[\varphi\right]-
\nonumber\\
&&-\pi\tilde{J}\sum_{\left\langle{\bf{r}},{\bf{r}}'\right\rangle}a^{2}\left({\bf{r}}\right)f\left({\bf{r}},{\bf{r}}'\right)\cos\left[\varphi({\bf{r}})-\varphi({\bf{r}}')\right]-
\nonumber\\
&&-2\pi\tilde{J}\sum_{\left\langle{\bf{r}},{\bf{r}}'\right\rangle}f\left({\bf{r}},{\bf{r}}'\right)\cos\left[\varphi({\bf{r}})-\varphi({\bf{r}}')\right].
\label{Equation_26}
\end{eqnarray}
Here, the free energy ${\cal{F}}_{\rm eff}\left[\varphi\right]$ is 
\begin{eqnarray}
{\cal{F}}_{\rm eff}\left[\varphi\right]=-\frac{1}{\beta}\sum_{{\bf{r}}}\ln\left[\frac{\sinh\left(\beta U\Omega_{{\bf{r}}}/2\right)}{\beta U\Omega_{{\bf{r}}}/2}\right].
\label{Equation_27}
\end{eqnarray}
The coefficients $f\left({\bf{r}},{\bf{r}}'\right)$ stands for the normal Gaussian smearing factors $f\left({\bf{r}},{\bf{r}}'\right)=\exp{\left[-a^{2}\left({\bf{r}}\right)/2-a^{2}\left({\bf{r}}'\right)/2\right]}/{2\pi}$. The parameters $\Omega_{{\bf{r}}}$ and $a^{2}\left({\bf{r}}\right)$ are inter-related by the minimization procedure. We have
\begin{eqnarray}
\Omega^{2}_{{\bf{r}}}=\frac{2\tilde{J}}{U}{\sum_{{\bf{r}}'}}'f\left({\bf{r}},{\bf{r}}'\right)\cos\left[\varphi({\bf{r}})-\varphi({\bf{r}}')\right],
\nonumber\\
a^{2}\left({\bf{r}}\right)=\frac{2}{U\beta\Omega^{2}_{{\bf{r}}}}\left[\frac{U\beta\Omega_{{\bf{r}}}}{4}\coth\left(\frac{U\beta\Omega_{{\bf{r}}}}{4}\right)-1\right],
\label{Equation_28}
\end{eqnarray}
where the prime on the sum in the first of equations in Eq.(\ref{Equation_28}) is over all ${\bf{r}}'$ nearest neighbors of ${\bf{r}}$. Furthermore, we have in Eq.(\ref{Equation_28}) an infinite number of coupled SC equations, which do not allow the solution in general. In contrast, at the low temperatures we can neglect thermal fluctuations of $\Omega_{{\bf{r}}}$ and $a\left({\bf{r}}\right)$ and we suppose that $\Omega_{{\bf{r}}}\equiv \Omega$ and $a\left({\bf{r}}\right)\equiv a$, i.e. the uniform solution. Hence, in the low-temperature limite, we have the classical Hamiltonian of the system given by
\begin{eqnarray}
{\cal{H}}_{\rm eff}\left[\varphi\right]=-\tilde{J}{\mathrm{g}}\sum_{\left\langle {\bf{r}},{\bf{r}}' \right\rangle}\cos\left[\varphi\left({\bf{r}}\right)-\varphi\left({\bf{r}}'\right)\right],
\label{Equation_29}
\end{eqnarray}
where ${\mathrm{g}}=\mathrm{g}_{0}\left(1-\ln{{\mathrm{g}}_{0}}\right)$ and, for the pair of parameters $(U, \tilde{J})$, ${\mathrm{g}}_{0}$ is given by the following equation 
\begin{eqnarray}
{\mathrm{g}}_{0}=\frac{T^{2}}{\tilde{J}U}\left(1-\frac{4\tilde{J}{\mathrm{g}}_{0}\ln{{\mathrm{g}}_{0}}}{T}\right)^{2}\cdot\tanh^{2}\left[\frac{\tilde{J}}{T}\sqrt{\frac{U{\mathrm{g}}_{0}}{\tilde{J}}}\right].
\label{Equation_30}
\nonumber\\
\ \ \ 
\end{eqnarray}
In fact, the classical phase Hamiltonian in Eq.(\ref{Equation_28}) is a usual $X-Y$ model Hamiltonian in 2D. Considering the bound-vortex-antivortex excitation state, the Hamiltonian in Eq.(\ref{Equation_29}) gives rise to a temperature-induced transition in the long wavelength limit of phase fluctuations. The equation for the BKT transition temperature is
\begin{eqnarray}
T_{\rm BKT}=\frac{\pi}{2}\tilde{J}\left(T_{\rm BKT}\right){\mathrm{g}}.
\label{Equation_31}
\end{eqnarray}
Furthermore, Eqs.(\ref{Equation_30}) and (\ref{Equation_31}), form a system of coupled SC equations for the parameter ${\mathrm{g}_{0}}$ and for the BKT transition temperature $T_{\rm BKT}$. We should take into account only the higher values of the parameter ${\mathrm{g}_{0}}$ and the temperatures $T_{\rm BKT}$  corresponding to it, for all other fixed values of the pair $(U,\tilde{J})$. Using again the fast convergent Newton's method, \cite{Powell} with the relative error of order $10^{-7}$, we find the solution $({\mathrm{g}_{0}},T_{\rm BKT})$ of the system. The corresponding values of the critical temperature $T_{\rm BKT}$ of the BKT superfluid transition  are given in Fig.~\ref{fig:Fig_3} for two different values of the $f$-band hopping $\tilde{t}$. For energy scale of $T_{\rm BKT}$ in the case of quasi-2D GaAs/AlGaAs QW structure geometry (see the Section \ref{sec:Section_4} for structure details) we find $T_{\rm BKT} \approx 12.06$ nK at $U=0.0188$ eV. For a In$_{0.08}$Ga$_{0.92}$As/GaAs coupled QW's the BKT transition temperature is slightly higher and $T_{\rm BKT}\approx 15.14$ nK, at the same values $U=0.0188$ eV. Note, that the recent experimental investigations for the probe of superfluidity in 2D Bose systems \cite{Julian} have proved the existance of the Landau critical velocity in the temperature range of order $90$ nK.
%
\section{Conclusions}

We have shown the existence of the BKT phase transition in 2D EFKM. Using the path integral approach and U(1) gauge representation of electron operator, we derived the general form of the action of the system. Then, we integrated out the fermionic and bosonic degrees of freedom, and we got the effective phase action (for the bosonic sector) and effective fermionic action (for the fermionic sector). Furthermore, the series expansion of the effective phase action up to second order in hopping amplitude, gives the form of the parameter $J$ of the phase-stiffness between n.n. excitonic pairs. Then, in the limit of low-temperatures, we have considered the ``vortex-antivortex" bound state modes, and we have reduced the quantum rotor model to the simple classical Hamiltonian description with the temperature dependent effective phase-stiffness parameter. Then, we calculated the BKT transition critical temperature $T_{\rm BKT}$ by solving numerically a system of coupled SC equations for the parameters $\mathrm{g}$ and $T_{\rm BKT}$. Thereby, we have shown that the BKT transition in the system of preformed excitonic pairs is directly related with the bound-vortex-antivortex type phase-configuration in the system, and the excitonic insulator state, with the order-parameter $\Delta$, is a necessary prerequisite for the realization of such a transition. In the frame of the same methods we used here, it still fundamental to answer the question, whether excitonic BEC transition temperature is coinciding with the critical temperature $T_{\rm BKT}$ of the excitonic superfluid transition, and for this, the quasi-2D system of excitons should be considered and the inter-layer exciton correlations would be properly included. We will discuss on that subject in the near future.

\section*{References}
%


\begin{thebibliography}{10}

\bibitem{Jerome} D. Jerome, T.M. Rice and W. Kohn, Phys. Rev. {\bf 158} 462, (1967).

\bibitem{Phan} V.N. Phan, H. Fehske and K.W. Becker, Europhysics letters {\bf 95} 17006, (2011).


\bibitem{Zenker} B. Zenker, D. Ihle, F.X. Bronold and H. Fehske, Phys. Rev. B, {\bf 85} 121102, (2012).


\bibitem{Seki}K. Seki, R. Eder and Y. Ohta, Phys. Rev. B, {\bf 84} 245106 (2011).

\bibitem{Lozovik_1} Yu.E. Lozovik, O.L. Berman and Tsvetus, V.G., JETP Lett. {\bf 66}, 332 (1997).
Phys. Rev. Lett. {\bf 82}, 871 (1999).

\bibitem{Lozovik_2} Yu.E. Lozovik and Sokolik A.A, V.G., JETP Lett. {\bf 84}, 61 (2007).
Phys. Rev. Lett. {\bf 82}, 871 (2007).

\bibitem{Hohenberg_1} D.S. Fisher and P.C. Hohenberg, Phys. Rev. B {\bf 37}, 4936 (1988).
Phys. Rev. Lett. {\bf 82}, 871 (2007).


\bibitem{Chang}C. Chang and R. Friedberg, Phys. Rev. B {\bf 51}, 1117 (1995).
Phys. Rev. Lett. {\bf 82}, 871 (2007).

\bibitem{Eisenstein} J.P. Eisenstein and A.H. MacDonald, Nature {\bf 432}, 691 (2004).

\bibitem{Clade}P.  Clad\'{e}, and C. Ryu and A. Ramanathan, K. Helmerson and W.D. Phillips, Phys. Rev. Lett. {\bf 102}, 170401 (2009).

\bibitem{Julian}R\'{e}mi Desbuquois and et al., Nature {\bf 8}, 645 (2012).

\bibitem{Chung}Ha Li-Chung and et al., Phys. Rev. Lett. {\bf 110}, 145302 (2013).

\bibitem{Kruger} P. Kr{\"u}ger, Z. Hadzibabic and J. Dalibard, Phys. Rev. Lett. {\bf 99}, 040402 (2007).

\bibitem{Bardeen} J. Bardeen, L.N. Cooper and J.R. Schrieffer, Phys. Rev. {\bf 108}, 1175 (1957).

\bibitem{Berezinskii} V.L. Berezinskii, Zh. Eksp. Teor. Fiz. (Sov. Phys. JETP) {\bf 61}, 1144 (1971).

\bibitem{Kosterlitz} J.M. Kosterlitz and D.J. Thouless, J. Phys. C {\bf 6}, 1181 (1973).

\bibitem{Mermin} N.D. Mermin and H. Wagner, Phys. Rev. Lett. {\bf 17}, 1133 (1966).

\bibitem{Hohenberg_2} P.C. Hohenberg, Phys. Rev. {\bf 158}, 383 (1967).

\bibitem{Landau} E.M. Lifshitz and L.P. Pitaevskii, Statistical Physics, Part II, {Pergamon, Oxford} (1981).

\bibitem{Hadzibabic}  C. Raman, M. K{\"o}hl, R. Onofrio and D.S. Durfee, C.E. Kuklewicz, Z. Hadzibabic and W. Ketterle, J. Phys. C,  {\bf 83}, 2502 (1999).

\bibitem{Sivan} U. Sivan, P.M.Solomon and H. Shtrikman, Phys. Rev. Lett.,  {\bf 68}, 1196 (1992).

\bibitem{Croxall} A.F. Croxall and et al., Phys. Rev. Lett.,  {\bf 101}, 246801 (2008).

\bibitem{Seamons} J.A. Seamons, C.P. Morath, J.L. Reno and M.P. Lilly, Phys. Rev. Lett. {\bf 102}, 026804 (2009).

\bibitem{Nelson_1} A. Perali, D. Neilson and A.R. Hamilton, Phys. Rev. Lett. {\bf 110}, 146803 (2013).

\bibitem{Negele} J.W. Negele and H. Orland, Quantum Many-Particle Systems, Addison-Wesley, Reading, MA (1988).

\bibitem{Kopec_1} T. K. Kope\'{c}, Phys. Rev. B, {\bf 70}, 054518 (2004).

\bibitem{Simanek} E. \v{S}im\'{a}nek, Phys. Rev. B {\bf 22}, 459 (1980).

\bibitem{Wood_Stroud} D.M. Wood and D. Stroud, Phys. Rev. B {\bf 25}, 1600 (1982).

\bibitem{Powell} M.J.D. Powell, A hybrid method for nonlinear equations,  in numerical methods of nonliear algebraic equations, Rabinowitz, P. ed., Gordon and Breach, New York (1970).

\bibitem{Kagan}  Yu. Kagan and et. al., Phys. Rev. A {\bf 61}, 043608 (2000).

\bibitem{Abramovich} M. Abramovitz and I. Stegun, Handbook of Mathematical Functions, Dover, New York, (1970).

\bibitem{Takahashi} Y. Takahashi and et al., J. Appl. Phys. {\bf 76}, 2299 (1994).

\bibitem{Szymanska} M.H. Szyma\'{n}ska and P.B. Littlewood, Phys. Rev. B {\bf 67}, 193305 (2003).

\bibitem{Kleinert_1} R.P. Feynman and H. Kleinert, Phys. Rev. A {\bf 34}, 5080 (1986).

\bibitem{Seunghwan} S. Kim and M.Y. Choi, Phys. Rev. B {\bf 41}, 111 (1990).




\end{thebibliography}
\end{document}